\documentstyle[12pt]{ioplppt}
\newcommand{\myvec}[1]{\mbox{\boldmath{$#1$}}}

\begin{document}
\title{First principle basis of the direct numerical simulation for
  turbulence of inert and reactive gases}
\author{
  S.Tsug\'e\\
  Tsukuba Advanced Research Alliance, University of Tsukuba,\\
  Tsukuba,305 Japan\ftnote{1}{present address : 790-3 Tohigashi, Tsukuba
    300-26 Japan}\ftnote{0}{e-mail address : shunt@tara.tsukuba.ac.jp}
  }

\begin{abstract}
An open question of whether phenomenological fluid equations to be
used for direct numerical simulation of turbulence are warranted on
`first principles' is addressed, and the problem is posed using
Klimontovich microscopic density to replace the Boltzmann function of
the classical statistical mechanics. For inert monatomic gases, it is
shown that all the gasdynamic equations, namely, the three
conservation equations plus the Navier-Stokes stress law and the
Fourier heat conduction law are retrieved as governing instantaneous
quantities, without having recourse to
any concepts of averaging or statistical equilibrium. For reactive
gases, however, the Arrhenius reaction rate law written in terms of the
fluctuating temperature is not justified, reflecting the fact that
this rate law hinges crucially on these concepts.
\end{abstract}


\section{Introduction}
Molecular approach to macroscopic behavior of gases by means of the
kinetic theory or nonequilibrium statistical mechanics has provided
firmer basis for variety of fields of phenomenologies.

Most important contributions among them include;
\begin{itemize}
\item[i)] derivation of the Navier-Stokes equation from Boltzmann's
  equation \cite{b1} \cite{b2},
\item[ii)] proof that the classical `equilibrium' thermodynamics
  remains exact to the first order in translational noneqilibrium,
  namely that it has the same region of validity as the Navier-
  Stokes equation, not the inviscid Euler equation \cite{b3}, and
\item[iii)] Arrhenius' rate law derived from the Boltzmann equation for
  simple classical molecules executing reactive collisions, where the
  activation energy is expressed by parameters of particle mechanics,
  namely, reduced mass and the threshold potential for inelastic
  collisions \cite{b4}.
\end{itemize}

Hence,those widely used tools of phenomenologies are warranted on
`first principles'.

If the flow is turbulent, however, it is not obvious that those
equations remain valid for instantaneous flow quantities that are both
stochastic and fractal \cite{b5}. In fact, as the matter of definition, the
Boltzmann function is the {\em mean} number density in the six-dimensional
phase(\(\mu\)-)space, so is subject to statistical
averaging. Therefore, the fluid velocity defined through a moment
integral of the Boltzmann function may not resolve small eddies or
fractal behavior.

Therefore, if the `first-principle' ground of the Navier-Stokes
equation to be used for the direct numerical simulation (DNS) is
asked, we have to address to {\em unaveraged\/} Boltzmann function and
to its governing equation that is free from statistical concept. Needs
for the approach along this line reflect the observation that
if we leave the binary (long range)correlation in the
Boltzmann equation, the Chapman-Enskog procedure applid to this equation
leads not to the `raw' Navier-Stokes equation, but to
Reynolds averaged one \cite{b6}.

Such a function known as the microscopic density is advocated by
Klimontovich in connection with plasma kinetic theory \cite{b7}. It
represents exact number density in the \(\mu\)- space, a deterministic
function by nature.

The equation governing the microscopic density for a neutral ideal gas
is derived in \cite{b8}. It is formulated as an equation of continuity in the
\(\mu\)- space, happens to be identical in form with Boltzmann's
equation with different implications. In what follows this equation will
be called master Boltzmann equation (MBE). Molecular background for
the fluid equations to be used for DNS should be a fluid version of the MBE.

The objective of this paper is to show how these fluid equations are
derived on first principles of molecular
dynamics for monatomic inert molecules, then how the situation
differs crucially for reactive molecules.

\section{A formalism on the microscopic density and the master
  Boltzmann equation}
Let the microscopic density be defined by the
following expression after Klimontovich \cite{b7}
\begin{equation}\label{eq1}
\underline{f}(z,t) \equiv \sum_{1 \le s \le N}\delta(z-z^{(s)}(t))
\end{equation}
where \(z \equiv (\myvec{x},\myvec{v})\) is a point in the \(\mu\)-
space, namely, a 6 dimensional space consisting of the physical- plus
molecular-velocity spaces, \(z^{(s)}(t)\) is locus of the \(s\)-th
molecule in this space, $\delta$ denotes the 6-D delta function and the
summation spans over total number of molecules N under consideration.
If $z^{(s)}(t)$ is given as the solution of Hamiltonian mechanics of
N-body system, formula (1) provides the most accurate expression for
the instantaneous number density in the $\mu$- space.

According to this `Klimontovich formalism' number density in the
physical space is given by
\begin{eqnarray}\label{eq2}
\underline{n}(\myvec{x},t) &\equiv& \int \underline{f}(z,t)d\myvec{v} \nonumber\\
&=&\sum_{1 \le s \le N}\delta(\myvec{x}-\myvec{x}^{(s)}(t))
\end{eqnarray}
In the same way fluid velocity \underline{\myvec{u}}(\myvec{x},t) is defined
by
\begin{eqnarray}\label{eq3}
\underline{n}\: \underline{\myvec{u}}(\myvec{x},t) &\equiv& \int \myvec{v}
\underline{f}(z,t)d\myvec{v} \nonumber\\
&=&\sum_{1 \le s \le N}\myvec{v}^{(s)}(t)\delta(\myvec{x}-\myvec{x}^{(s)}(t))
\end{eqnarray}
These expressions are to be compared with those of the Boltzmann
formalism of the classical kinetic theory of gases;
\begin{equation}\label{eq4}
n=\int f d \myvec{v}
\end{equation}
\begin{equation}\label{eq5}
\overline{\underline{n}\: \underline{\myvec{u}}}= \int \myvec{v}f d\myvec{v}
\end{equation}
where $f$ is the Boltzmann function defined by
\begin{equation}\label{eq6}
f = \underline{\overline{f}}
\end{equation}
In this equation overbar( $\bar{ }$ ) denotes the standard `average', namely
ensumble average over repeated experiments, or equivalently, an
average of hotwire output over a time long enough to smooth out
fractals, yet short enough to resolve macroscopic unsteadiness such as
periodic shedding of K\'arm\'an vortices.

We note that there are three different levels of description of the
fluid velocity: Obviously expression (\ref{eq3}) is the most exact one
thinkable if $[\myvec{x}^{(s)}(t), \myvec{v}^{(s)}(t)]$ are solved
from Hamiltonian mechanics of the N-particle system. Therefore it
is a deterministic quantity even if the flow is turbulent.

This quantity, however, does not correspond to the fluid velocity as
observed by macroscopic measurements such as hotwire anemometry. Such
velocity is represented by the one in which the microscopic density
$\underline{f}$ is replaced with its spacial average
\begin{equation}\label{eq7}
\widetilde{\underline{f}} \equiv \frac{1}{\Delta \myvec{X}} \int_{\Delta
  \myvec{X}}d \myvec{x} \underline{f}
\end{equation}
where the integration is taken over a small, yet macroscopic volume
$\Delta \myvec{X} (= \Delta x_{1}\Delta x_{2}\Delta x_{3})$ such that
\begin{equation}\label{eq8}
n \Delta \myvec{X} \gg 1
\end{equation}
This is where stochastic facet of turbulence makes its appearance,
therefore turbulent fluid velocity defined in terms of
$\widetilde{\underline{f}}$ shows both stochastic and fractal behavior.

The third level of description of fluid velocity is
expression(\ref{eq5}) that is defined by the Boltzmann function, where
the average is taken over a long time or a large number of
repeated experiments so that both stochastic and fractal features are
averaged out. The third level is what is called Reynolds-average
formalism or the Boltzmann formalism, and will not be discussed here.

It is shown that equation for $\underline{f}$ of the first level of
description has been derived as the equation of continuity in the
$\mu$-space. For the usual monatomic gases it is given in the form of
Boltzmann's equation without invoking any statistical concept \cite{b8}
\begin{equation}\label{eq9}
B(\underline{f}) \equiv (\frac{\partial}{\partial t} + \myvec{v} \cdot
\frac{\partial}{\partial \myvec{x}}) \underline{f} - J
[\underline{f}\:\underline{\widehat{f}}\:]= 0
\end{equation}
with $J$ denoting the classical Boltzmann collision integral defined by
\begin{equation}\label{eq10}
J [\underline{f}\:\underline{\widehat{f}}\:] \equiv
\int[\underline{f}(z^{\prime})\underline{f}(\widehat{z}^{\prime})-
\underline{f}(z)\underline{f}(\widehat{z})]_{\myvec{x}=\widehat{\myvec{x}}} d \widehat{\myvec{x}} d
\widehat{\myvec{v}}
\end{equation}
where the prime $( ^{\prime})$ denotes a phase-space point of a particle
that is supposed to reach the unprimed point after collision,
and integration $\int d \widehat{\myvec{x}}$ spans over the physical space of
particle $\widehat{z}$ supposed to collide within unit of time.

The master-Boltzmann equation(MBE), namely, Eq, (\ref{eq9}) for the microscopic density, serves the most
precise description of collective behavior of the ensemble of the
$N$-particle system. It can describe thermal agitations that are not
tangible to macroscopic sensors. See ref.\cite{b8} for derivation of the
Landau-Lifshitz stress/heat flux density fluctuation formulae with
even finer temporal resolvability than the original derivation \cite{b9}.

It should be stressed that MBE holds for
$\underline{f}$, not for $\widetilde{\underline{f}}$ unless
$$\widetilde{\underline{f}\:\underline{f}}=\widetilde{\underline{f}}\:\widetilde{\underline{f}}$$
a condition that lacks immediate justification is met.

\section{Conservation equations as fluid moments of the master Boltzmann equation}
Here we address to equations governing instantaneous fluid quantities
by taking fluid moments of the MBE. A parallelism to the classical
procedure by Chapman-Enskog or Grad \cite{b10} prevails also here except that
the microscopic density replaces the Boltzmann function, namely that
exact delta function representation replaces Hermite polynomial
expansion around the Maxwellian distribution.

General expression for the moment equations of the MBE
(\ref{eq9}) takes the usual form as follows,
\begin{eqnarray}\label{eq11}
\left.
\begin{array}{cc}
&\displaystyle\frac{\partial}{\partial t} \int M \underline{f} d \myvec{v}+
\frac{\partial}{\partial x_{r}} \int v_{r} M \underline{f} d
\myvec{v}=I(M) \\
\mbox{with}& \\
&\displaystyle I(M)\equiv \frac 12 \int [[M]] \underline{f}\: \underline{\widehat{f}} d
\widehat{\myvec{x}} d \myvec{v} d \widehat{\myvec{v}} \\
&[[M]]\equiv M^{\prime}+ \widehat{M^{\prime}}-M- \widehat{M}
\end{array}\right\}
\end{eqnarray}
where $M$ stands for one of the following 13 moment functions;
\begin{equation}\label{eq12}
m,m v_{j},\frac m2 v^{2}_{k},m(w_{j} w_{l}- \frac13 \delta_{j l}
w^{2}_{k}),\frac m2 w^{2}_{k} w_{j} \  (m;\mbox{mass of a molecule})
\end{equation}
with
\begin{equation}\label{eq13}
w_{j} \equiv v_{j}- \underline{u}_{j}
\end{equation}
For the first three $M$'s that are summational invariants, $I(M)$ is
shown to vanish, and three conservation equations, same in form with
the classical theory, result;
\begin{equation}\label{eq14}
\frac{\partial \underline{\rho}}{\partial t}+
\frac{\partial(\underline{\rho}\underline{u}_{j})}{\partial x_{j}}=0
\end{equation}
\begin{equation}\label{eq15}
\frac{\partial}{\partial
  t} (\underline{\rho}\underline{u}_{j})+\frac{\partial}{\partial
  x_{r}}(\underline{\rho}\underline{u}_{j}\underline{u}_{r}+ \delta_{jr} \underline{p}+ \underline{p}_{jr})=0
\end{equation}
\begin{equation}\label{eq16}
\frac{\partial}{\partial
  t}[\underline{\rho}(\underline{e}+\frac{\underline{u}^{2}_{r}}{2})]+\frac{\partial}
{\partial x_{r}}[\underline{\rho}\underline{u}_{r}(\underline{h}+\frac{\underline{u}^{2}_{k}}{2})+\underline{q}_{r}+\underline{u}_{k}\underline{p}_{kr}]=0
\end{equation}
Fluid quantities to appear in these equations differ from those of
Chapman-Enskog theory in that there are no statistical concepts
intervening. (See, for instance, Eqs, (\ref{eq2}) and (\ref{eq3}), in
contrast with (\ref{eq4}) and (\ref{eq5}).) For the rest of quantities the following definitions are employed;
\begin{eqnarray}\label{eq17}
\underline{p} \equiv \frac m3 \int w^{2}_{r} \underline{f} d \myvec{v} \equiv \frac
m3 \sum_{1 \le s \le N}(v^{(s)}_{r} - \underline{u}_{r})^{2} \delta (\myvec{x}-\myvec{x}^{(s)}(t))
\end{eqnarray}
\begin{eqnarray}\label{eq18}
\left.
\begin{array}{cc}
&\displaystyle\frac 32 k \underline{T} \equiv \frac {1}{\underline{n}} \int \frac m2  w^{2}_{r}
\underline{f} d \myvec{v}= \frac 32 \frac{\underline{p}}{\underline{n}} \
(k;\mbox{Boltzmann's \ constant)} \\
\vspace{-5mm}\\
&\displaystyle\underline{e}=\frac 32 \frac{k}{m}\underline{T} \\
\vspace{-5mm}\\
&\displaystyle\underline{h}=\frac 52 \frac{k}{m}\underline{T}
\end{array}
\right\}
\end{eqnarray}
\begin{eqnarray}\label{eq19}
\underline{p}_{j l} &\equiv& m \int w_{j} w_{l} \underline{f} d \myvec{v} -
\frac{\delta_{j l}}{3} \underline{p} \nonumber\\
&=& m \sum_{1 \le s \le N}
[(v_{j}^{(s)}-\underline{u}_{j})(v_{l}^{(s)}-\underline{u}_{l})-\frac{\delta_{j l}}{3}(v_{k}^{(s)}-\underline{u}_{k}^{(s)})^{2}]
  \delta(\myvec{x}-\myvec{x}^{(s)})
\end{eqnarray}
\begin{eqnarray}\label{eq20}
\underline{q}_{j} &\equiv& \frac m2 \int w^{2}_{r} w_{j} \underline{f}
d\myvec{v}\hspace{67mm}\nonumber\\
&=& \frac m2 \sum_{1 \le s \le N}
(v_{r}^{(s)}-\underline{u}_{r})^{2}(v^{(s)}_{j}-\underline{u}_{j})\delta(\myvec{x}-\myvec{x}^{(s)})
\end{eqnarray}
where $\underline{p},\underline{e},\underline{h},\underline{T},\underline{p}_{j l}$ and $\underline{q}_{j}$ denote,
respectively, pressure, specific internal energy and enthalpy,
temperature, viscous stress tensor, and heat flux density. All of
these quantities correspond to those to be computed from DNS with
finest (the first) level of resolvability, much finer than those
detected by any macroscopic measurements that belong to the second
level of description.

\section{Transport relationships for viscous stress and heat flux
  density}

Conservation equations (\ref{eq14}) through (\ref{eq16}) are not closed; two more relationships that give $\underline{p}_{j l}$ and $\underline{q}_{j}$ in terms of other quantities are needed. Eq. (11) for
$$M; m(w_{j}w_{l}-\frac13 \delta_{j l}w_{k}^{2}),\frac m2 w^{2}_{k} w_{j}$$
will provide such relationships.

For this purpose somewhat tedious calculations of
\begin{equation}\label{eq21}
  I_{j l} \equiv I[m(w_{j} w_{l}- \frac13 \delta_{j l} w^{2}_{k})]
\end{equation}
\begin{equation}\label{eq22}
I_{j} \equiv I[ \frac m2 w_{k}^{2} w_{j}]
\end{equation}
will be carried out. Substituting (\ref{eq1}) into (\ref{eq11}) we have
\begin{equation}\label{eq23}
I(M)= \frac 12 \int [[M]] \sum_{s,\hat{s}}
\delta(\myvec{x}-\myvec{x}^{(s)})\delta(\widehat{\myvec{x}}-\myvec{x}^{(\hat{s})})
\delta(\myvec{v}-\myvec{v}^{(s)})\delta(\widehat{\myvec{v}}-\myvec{v}^{(\hat{s})})
d \widehat{\myvec{x}} d \myvec{v} d \widehat{\myvec{v}}
\end{equation}
It is more convenient to transform velocity variables to relative and
center-of-gravity ones as
\begin{eqnarray}\label{eq24}
\left.
\begin{array}{cc}
&\displaystyle\myvec{V}=\widehat{\myvec{v}}- \myvec{v}, \ \myvec{U}= \frac 12
(\myvec{v}+\widehat{\myvec{v}})\\
&\displaystyle\frac{\partial(\myvec{U},\myvec{V})}{\partial(\myvec{v},\widehat{\myvec{v}})}=1
\end{array}
\right\}
\end{eqnarray}
Note also that
\begin{equation}\label{eq25}
\int d \widehat{\myvec{x}}=\int V d\Omega
\end{equation}
where $d\Omega=d\Omega(s,\hat{s})$ is the differential cross-section
for a molecular encounter of a pair of particles $(s,\hat{s})$ such that
\begin{equation}\label{eq27}
V d\Omega=Vbdbd \epsilon = B(V,\theta) d \theta d \epsilon
\end{equation}
with $b,\epsilon$ and $\theta$ denoting impact parameter, azymuthal
angle, and bisect angle in the physical space between approaching and leaving asymptote loci
of $\hat{s}$ relative to $s$.
For `hard' repulsive intermolecular force $F=aR^{-n}(n \geq 3)$, elementary calculation of two-body mechanics gives
\begin{equation}\label{eq28}
B(\theta,V)=(\frac{2a}{m})^{\frac{2}{n-1}}V^{\frac{(n-5)}{(n-1)}}\beta \frac{d\beta}{d\theta}
\end{equation}
where $\beta$ is a scaled impact parameter
$$ \beta \equiv (\frac{m}{2a})^{\frac{1}{n-1}}V^{\frac{2}{n-1}}b$$
introduced to eliminate $V$-dependence of $\theta(\beta)$;
\begin{eqnarray}\label{eq29}
\theta(\beta)= \int_{0}^{\tau_{0}} \frac{d
  \tau}{[1-\tau^{2}-\frac{2}{n-1}(\frac{\tau}{\beta})^{n-1}]^{\frac
    12}}\nonumber\\
1-\tau_{0}^{2}-\frac{2}{n-1}(\frac{\tau_{0}}{\beta})^{n-1}=0
\end{eqnarray}
In what follows, unless stated otherwise, we will employ Maxwellian
molecules ($n=5$). This is the particular case in which $B(\theta,V)$
is independent of $V$, yet representing a realistic model
for usual neutral gas molecules.(See for example ref.\cite{b10}.)

With this specification for the molecular property,
calculation of $I(M)$ proceeds as follows :
\begin{eqnarray}\label{eq30}
I(M) &=&\int B d \theta d \epsilon \sum_{s,\hat{s}}
\delta(\myvec{x}-\myvec{x}^{(s)})\delta(\myvec{x}-\myvec{x}^{(\hat{s})})
\nonumber\\ && \times\int
\frac12[[M]]\delta(\myvec{U}-\frac{\myvec{w}^{(s)}+\myvec{w}^{(\hat{s})}}{2})
\delta(\myvec{V}-\myvec{w}^{(\hat{s})}+\myvec{w}^{(s)})d\myvec{U}d\myvec{V}
\nonumber \\ &=&
\sum_{s,\hat{s}}\delta(\myvec{x}-\myvec{x}^{(s)})
\delta(\myvec{x}-\myvec{x}^{(\hat{s})})\int_{0}^{2\pi}d\epsilon
\int_{0}^{\frac{\pi}{2} }d\theta B(\theta) \nonumber \\ &&
\times K(\frac{\myvec{w}^{(s)}+\myvec{w}^{(\hat{s})}}{2},
\myvec{w}^{(\hat{s})}-\myvec{w}^{(s)}, \theta, \epsilon)
\end{eqnarray}
where we have defined
\begin{equation}\label{eq31}
K(\myvec{U},\myvec{V},\theta,\epsilon) \equiv \frac12 [[M]]
\end{equation}
The actual forms of $K$ together with the integrals above for
$M=mw_{j}w_{l}$ and $mw_{j}\frac{w^{2}}{2}$ have been given by
Grad \cite{b10}, yielding
\begin{eqnarray}\label{eq32}
\int K(\myvec{U},\myvec{V},\theta,\epsilon)B(\theta)d\theta d\epsilon =
\left\{
\begin{array}{c}
\displaystyle -3mB_{1}(V_{j}V_{l}-\frac{\delta_{j l}}{3}V^{2})\\
\displaystyle-\frac32 mB_{1}(V_{j}V_{k}U_{k}-\frac{V^{2}}{3}U_{j})
\end{array}\right.
\end{eqnarray}
respectively, where $B_{1}$ is a constant defined by the following integral
\begin{equation}\label{eq33}
B_{1}=\pi \int^{\frac{\pi}{2}}_{0} B(\theta)\sin^{2}\theta
\cos^{2}\theta d\theta
\end{equation}
Eq.(\ref{eq32}) substituted into (\ref{eq30}) with
$(\myvec{U},\myvec{V})$ replaced by
$(\frac12
(\myvec{w}^{(s)}+\myvec{w}^{(\hat{s})}),\myvec{w}^{(\hat{s})}-\myvec{w}^{(s)})$
leads to
\begin{eqnarray}\label{eq34}
I_{j l}&=&-3mB_{1}\sum_{s,\hat{s}}\delta(\myvec{x}-\myvec{x}^{(s)})\delta(\myvec{x}-\myvec{x}^{(\hat{s})})\nonumber \\
&&
\times [(w_{j}^{(s)}-w_{j}^{(\hat{s})})(w_{l}^{(s)}-w_{l}^{(\hat{s})})
-\frac{\delta_{j l}}{3}(w_{k}^{(s)}-w_{k}^{(\hat{s})})^{2}]\nonumber\\
&=& -3mB_{1}[\sum_{s}\delta(\myvec{x}-\myvec{x}^{(s)})
\sum_{\hat{s}}\delta(\myvec{x}-\myvec{x}^{(\hat{s})})
(w_{j}^{(\hat{s})}w_{l}^{(\hat{s})}-\frac{\delta_{j l}}{3}w_{k}^{(\hat{s})^2})
\nonumber \\
&& +\sum_{\hat{s}}\delta(\myvec{x}-\myvec{x}^{(\hat{s})})
\sum_{s}\delta(\myvec{x}-\myvec{x}^{(s)})
(w_{j}^{(s)}w_{l}^{(s)}-\frac{\delta_{j l}}{3}w_{k}^{(s)^2})\nonumber \\
&& -2\sum_{s}\delta(\myvec{x}-\myvec{x}^{(s)})w_{j}^{(s)}
\sum_{\hat{s}}\delta(\myvec{x}-\myvec{x}^{(\hat{s})})w_{l}^{(\hat{s})}
\nonumber \\
&& +\frac 23 \delta_{j l}\sum_{s}\delta(\myvec{x}-\myvec{x}^{(s)})w_{k}^{(s)}
\sum_{\hat{s}}\delta(\myvec{x}-\myvec{x}^{(\hat{s})})w_{k}^{(\hat{s})}]
\nonumber \\
&=& -6B_{1}\underline{n}\underline{p}_{j l}
\end{eqnarray}

\begin{eqnarray}\label{eq34.5}
I_{j}&=&-2mB_{1}\sum_{s,\hat{s}}\delta(\myvec{x}-\myvec{x}^{(s)})\delta(\myvec{x}-\myvec{x}^{(\hat{s})})[\frac12
(w_{j}^{(s)}w_{k}^{(s)^{2}}+
w_{j}^{(\hat{s})}w_{k}^{(\hat{s})^{2}})\nonumber \\
&&-(w_{j}^{(s)}w_{k}^{(\hat{s})^{2}}+w_{j}^{(\hat{s})}w_{k}^{(s)^{2}})+\frac12
(w_{k}^{(s)}w_{j}^{(\hat{s})}w_{k}^{(\hat{s})}+
w_{j}^{(s)}w_{k}^{(s)}w_{k}^{(\hat{s})})] \nonumber \\
&=&-4B_{1}\underline{n}\underline{q}_{j}
\end{eqnarray}
where definitions (\ref{eq13}) for $w_{j}$ and
(\ref{eq19}), (\ref{eq20}) for
$\underline{p}_{j l}$ and $\underline{q}_{j}$, respectively, have been made use of. Lines of lengthy deduction of algebra have been suppressed in deriving (\ref{eq34.5}).

The lefthand side of the moment equation (\ref{eq11}) for the stress
tensor and the heat flux density are
calculated using only fluid quantities as defined here, namely, those
with no statistical average involved, yet leading to the same
expression as the 13-moment method (Eq. (\ref{eq5},
\ref{eq17}),\cite{b10}) as;
\begin{equation}\label{eq35}
\underline{p}(\frac{\partial \underline{u}_{j}}{\partial x_{l}}+\frac{\partial
  \underline{u}_{l}}{\partial x_{j}}-\frac 23 \delta_{j l}\frac{\partial
  \underline{u}_{k}}{\partial x_{k}})+O(\underline{p}_{j l},\underline{q}_{j})
\end{equation}
and
\begin{equation}\label{eq35.5}
\frac52 \frac km \underline{p} \frac{\partial \underline{T}
}{\partial x_{i}}+O(\underline{q}_{i},\underline{p}_{il})
\end{equation}
respectively. Equating expressions (\ref{eq34}) and (\ref{eq35}) we have
\begin{eqnarray}\label{eq36}
\left.
\begin{array}{crcl}
&\underline{p}_{j l}&=&\displaystyle - \underline{\mu}(\frac{\partial \underline{u}_{j}}{\partial
  x_{l}}+\frac{\partial \underline{u}_{l}}{\partial x_{j}}-\frac 23
\delta_{j l}\frac{\partial \underline{u}_{k}}{\partial x_{k}})\\
\mbox{with}\hspace{15mm}& \\
&\underline{\mu} &\equiv& \displaystyle\frac{k \underline{T}}{6B_{1}} \hspace{2mm} \mbox{(viscosity coefficient)}
\end{array}\right\}
\end{eqnarray}
likewise, from (\ref{eq34.5}) and (\ref{eq35.5})
\begin{eqnarray}\label{eq37}
\left.
\begin{array}{crcl}
& \underline{q}_{j}&=&\displaystyle -\lambda \frac{\partial \underline{T}}{\partial x_{j}}
\\
\mbox{with}\hspace{10mm}& \\
& \lambda&=&\displaystyle\frac{5(k/m)}{8B_{1}}k\underline{T} \hspace{5mm} \mbox{(heat
  conductivity coefficient)}
\end{array}\right \}
\end{eqnarray}
both exact to the first
order smallness of $\underline{p}_{j l}$ and $\underline{q}_{j}$.

Formulae (\ref{eq36}) and (\ref{eq37}) are nothing but the
Navier-Stokes and the Fourier laws for
fluctuating fluid quantities and viscosity/heat conductivity
coefficients that also fluctuate with temperature.

It is rather surprising that these transport relationships that have
been derived originally as the first-order deviation from the
equilibrium(Maxwellian) distribution prove to be correct even without
any statistical concept. In fact, formulae (\ref{eq36}) and
(\ref{eq37}) are written in terms of collective coordinates of
N-particle Hamiltonian mechanics alone. Their validity is subject to the only
 assumption that both $\underline{p}_{jl}$ and $\underline{q}_{j}$ are of the first order
smallness compared with those quantities representing first terms of (35)
and (36), rspectively.

All the quantities on the r.h.s. of (\ref{eq36}) and (\ref{eq37})
correspond to hotwire
or thermocouple outputs that are stochastic and fractal. Seeming difficulty
of partial differential representations of such fractal
(therefore not differentiable) quantities are remedied by the
experimental evidence
that the fractal structure does not persist to infinitesimal region,
but ceases to operate at Kolmogorov length \cite{b11}. Real
difficulties arise when the differentials are replaced with
finite differences for computational purposes, in which case a step size
be chosen smaller than the Kolmogorov length (the high Reynolds number
difficulty).

\section{Chemical reaction rate}
Chemical reactions of the following type
$$
A+B \to C+D
$$
that are caused by inelastic binary collisions
have reaction rates obeying Arrhenius' low
\begin{equation}\label{eq39}
W \sim e^{-E/RT}
\end{equation}
Present \cite{b4} has derived this rate law using a simple collision model,
where the activation energy $E$ is expressed in terms of molecular
parameters, namcly, threshold potential $\epsilon$ for inelastic collisions
and reduced mass $m_{AB}$ as
\begin{equation}\label{eq40}
E=\frac{\epsilon}{2 m_{AB}}
\end{equation}

It is an open question that how the rate law (\ref{eq39}) is modified
if microscopic density is used instead of the Maxwellian distributin
in calculating the reaction rate;
\begin{equation}\label{eq41}
\underline{W}=\int_{\mbox{inel.}} \underline{f}_{A}\:\underline{f}_{B} d \myvec{v}_{A} d
\myvec{v}_{B} Vd \Omega_{AB}
\end{equation}
First we note that the following relationship holds independently of
the intermolecular force law
\begin{eqnarray}\label{eq42}
\int_{\mbox{inel.}}d\Omega_{AB}&=&\int_{0}^{2\pi}d\epsilon
\int_{0}^{b^{\dagger}} bdb \nonumber \\
&=&\pi b^{\dagger 2}
\end{eqnarray}
where $b^{\dagger}$ denotes the maximum impact parameter for a
reactive collision to occur. The value of $b^{\dagger}$ can be
obtained from the conservation laws
of angular momentum and energy of two-body mechanics,
$$\left.
\begin{array}{rcll}
R^{2}\dot{\theta}&=&bV &:\; \mbox{const} \\
\displaystyle(\frac{m_{AB}}{2})(\dot{R}^{2}+R^{2}\dot{\theta}^{2})+\Omega(R)
&=&\displaystyle(\frac{m_{AB}}{2})V^{2}&:\; \mbox{const}
\end{array}\right\}
$$
where $(R(t),\theta(t))$ is the relative particle trajectory,
$\Omega(R)$ is the intermolecular potential as depicted in Fig.1.\\

Eliminating $\dot{\theta}$ and solving for the critical impact parameter
$b=b^{\dagger}$ corrresponding to the trajectory of the closest
elastic encounter that is tangential to the ridge line
$(dR/d\theta =0 \ \mbox{at} \ R=r_{0},\Omega(r_{0})=\epsilon)$
of the potential `crater', we have
\begin{equation}\label{eq43}
\frac{b^{\dagger}}{r_{0}}=(1-\frac{2\epsilon}{m_{AB}V^{2}})^{\frac12}
\end{equation}
Substitution of Eqs.(\ref{eq42}) and (\ref{eq43}) into (\ref{eq41})
gives
\begin{eqnarray}\label{eq44}
\left.
\begin{array}{rcl}
\underline{W}&=&\displaystyle\pi
r_{0}^{2}\int_{V>V_{cr}}d\myvec{V}V(1-\frac{V_{cr}^{2}}{V^{2}})\int
\underline{f}_{A}\: \underline{f}_{B}d\myvec{U}\\
\vspace{-2mm}\\
V_{cr}&\equiv&\displaystyle
(\frac{2\epsilon}{m_{AB}})^{\frac12}=[2E(M_{A}^{-1}+M_{B}^{-1})]^{\frac12},
\ (M ; \mbox{molarweight})
\end{array}
\right\}
\end{eqnarray}
valid for any force laws governing elastic part of collisions.
Employing definition (\ref{eq1}) for the microscopic density, rate of
reaction (\ref{eq44}) is written in an alternative form
\begin{eqnarray}\label{eq45}
&&\underline{W}=\pi
r_{0}^{2}\sum_{s,\hat{s}}\delta(\myvec{x}-\myvec{x}_{A}^{(s)})\delta(\myvec{x}-\myvec{x}_{B}^{(\hat{s})})
\nonumber \\
&&\hspace{20mm} \int_{V>V_{cr}}\delta(\myvec{V}-\myvec{v}_{B}^{(\hat{s})}+\myvec{v}_{A}^{(s)})V(1-\frac{V_{cr}^{2}}{V^{2}})d\myvec{V}
\end{eqnarray}
In calculating the integral, we make use of the following formula
\begin{eqnarray}\label{eq46}
&&\int_{V>V_{cr}}F(V)\delta(\myvec{V}-\widehat{\myvec{V}})
d\myvec{V}=F(\widehat{V})\int_{V>V_{cr}}
\delta(\myvec{V}-\widehat{\myvec{V}})d\myvec{V} \nonumber \\
&&\hspace{10mm} =F(\widehat{V})\int dV_{x}dV_{y} \delta(V_{x}-\widehat{V}_{x})
\delta(V_{y}-\widehat{V}_{y})S(\widehat{V}_{z};
\sqrt{V_{cr}^{2}-V_{x}^{2}-V_{y}^{2}}) \nonumber \\
&&\hspace{10mm} =F(\widehat{V})S(\widehat{V}_{z};
\sqrt{V_{cr}^{2}-\widehat{V}_{x}^{2}-\widehat{V}_{y}^{2}}) \nonumber \\
&&\hspace{10mm} =F(\widehat{V})S(\widehat{V};V_{cr})
\end{eqnarray}
where $S$ is the step function defined by
\begin{eqnarray}\label{eq47}
S(X;Y)=\left\{
\begin{array}{cc}
1, \  & \mbox{if} \ X>Y \\
0, \  & \mbox{otherwise}
\end{array}
\right.
\end{eqnarray}
Then Eq. (\ref{eq45}) reads
\begin{eqnarray}\label{eq48}
\left.
\begin{array}{rl}
\underline{W}=& \pi r_{0}^{2}\displaystyle{\sum_{s,\hat{s}}} \delta
(\myvec{x}-\myvec{x}_{A}^{(s)})
\delta(\myvec{x}-\myvec{x}_{B}^{(\hat{s})})
S(V_{AB};V_{cr})V_{AB}(1-\frac{V_{cr}^{2}}{V_{AB}^{2}}) \\
\mbox{with} & \\
\myvec{V}_{AB}\equiv& \myvec{v}_{B}^{(\hat{s})}-\myvec{v}_{A}^{(s)}
\end{array}
\right\}
\end{eqnarray}
This expression is the ultimate form of the instantaneous reaction
rate after the Klimontovich formalism;
further deduction is not possible unless any assumption is made on
$V_{AB}$; its average $\widetilde{V}_{AB}=(8kT/ \pi
  m_{AB})^{1/2}$ for instance. Then we have
\begin{eqnarray}\label{eq49}
\widetilde{\underline{W}}=\alpha
\left\{
\begin{array}{cl}
\displaystyle 1-\frac{\pi}{4}(\frac{\epsilon}{kT})& , \mbox{for} \hspace{5mm}
\displaystyle kT>(\frac{\pi}{4})\epsilon \\
\vspace{-2mm}\\
0& ,  \mbox{otherwise}
\end{array}
\right.
\end{eqnarray}
where
$$
\alpha \equiv \pi r_{0}^{2} \underline{n}_{A}\underline{n}_{B} (\frac{8kT}{\pi
  m_{AB}})^{\frac12}
$$
is the frequency factor.

Another model to make formula (\ref{eq48}) tractable would be to take
the average of the following type;
\begin{equation}\label{eq50}
\widetilde{\underline{W}} \equiv \int \underline{W} F(V_{AB})d\myvec{V}_{AB}
\end{equation}
where
$$
F(V_{AB}) \equiv \frac{1}{(2\pi
  c_{AB}^{2})^{\frac32}}\exp(-\frac{V_{AB}^{2}}{2c_{AB}^{2}}),\hspace{5mm}
(c_{AB}^{2} =\frac{kT}{m_{AB}})
$$
is the probability density of the relative velocity distribution in
statistical equilibrium. Then the classical Arrhenius rate
law written in terms of the average temperature
\begin{equation}\label{eq51}
\widetilde{\underline{W}}=\alpha e^{- \epsilon /kT}
\end{equation}
is recovered as it should be.

In Fig.2 is shown the comparison of the two average reaction rates
(\ref{eq49}) and (\ref{eq51}) plotted against the inverse Zeldovich
number $(\epsilon /kT)^{-1}$. The former which might be called the
ignition temperature
model predicts the ignition temperature $T_{i}$ as
$$
kT_{i}=(\frac{\pi}{4})\epsilon
$$
in somewhat overestimating manner of reality.

\section{Conclusions}

The open question of whether the fluid equations as used for the
direct numerical simulation of turbulence, namely, those written
in terms of fluctuating variables have `first principle' basis is
looked into by the molecular approach using the microscopic
density. On this Klimontovich formalism as contrast to the Boltzmann
formalism of the classical statistical mechanics, three conservation
equations, the Navier-Stokes viscous-stress law and the Fourier heat
conduction law prove to be valid with much finer spatial
resolvability, and without any concepts of averaging or statistical
equilibrium.

For reactive gases, however, the parallelism breaks down in that the
Arrhenius reaction rate law depends crucially on these concepts,
therefore that this rate law written by the fluctuating temperature is
not justified.

\newpage

\section*{Figure captions}

Fig.1 \ The intermolecular potential with threshold \ $\epsilon$ \ for
reactive collisions \\
Fig.2 \ Arrhenius-and ignition-temperature models

\end{document}